\newif\iffigs\figsfalse
\figstrue

\documentstyle[11pt]{article}
\iffigs
  \input epsf
\else
\fi

\begin{document}



\def\ga{\alpha}
\def\gb{\beta}
\def\gc{\gamma}
\def\gcp{\gamma^\prime}
\def\gd{\delta}
\def\gep{\epsilon}
\def\gl{\lambda}
\def\gL{\Lambda}
\def\gk{\kappa}
\def\go{\omega}
\def\gp{\phi}
\def\gs{\sigma}
\def\gt{\theta}
\def\gC{\Gamma}
\def\gD{\Delta}
\def\gO{\Omega}
\def\gT{\Theta}


\def\be{\begin{equation}}
\def\ee{\end{equation}}
\def\ba{\begin{array}}
\def\ea{\end{array}}
\def\bea{\begin{eqnarray}}
\def\eea{\end{eqnarray}}
\def\bes{\begin{eqnarray*}}
\def\ees{\end{eqnarray*}}
\def\bsea{\begin{subeqnarray}}
\def\esea{\end{subeqnarray}}
\def\btab{\begin{tabular}}
\def\etab{\end{tabular}}
\def\lra{\longrightarrow}
\def\lms{\longmapsto}
\def\ra{\rightarrow}
\def\ms{\mapsto}
\def\pa{\partial}
\def\ll{\parallel}


\newcounter{subequation}[equation]

\makeatletter
\expandafter\let\expandafter\reset@font\csname reset@font\endcsname
\newenvironment{subeqnarray}
  {\def\@eqnnum\stepcounter##1{\stepcounter{subequation}{\reset@font\rm
      (\theequation\alph{subequation})}}\eqnarray}%
  {\endeqnarray\stepcounter{equation}}
\makeatother

\newtheorem{lemma}{Lemma}
\newtheorem{satz}{Satz}
\newtheorem{theorem}{Theorem}
\newtheorem{cor}{Corollary}
\newtheorem{define}{Definition}


\newcommand{\N}{\mbox{I\hspace{-.4ex}N}}
\newcommand{\C}{\mbox{$\,${\sf I}\hspace{-1.2ex}{\bf C}}}
\newcommand{\Cs}{\mbox{$\,${\sf I}\hspace{-1.2ex}C}}
\newcommand{\Z}{\mbox{{\sf Z}\hspace{-1ex}{\sf Z}}}
\newcommand{\R}{\mbox{\rm I\hspace{-.4ex}R}}
\newcommand{\1}{\mbox{1\hspace{-.6ex}1}}

\def\la{\wi{\mbox{Tr} J^2}}
\def\cp{c^\prime}
\def\rum{r_{\mu_{\rm min}}}
\newcommand{\me}{\mbox{e}}
\newcommand{\wum}{\left(1 + \frac{\tau^2}{\omega^2}\right)^{-\frac{1}{2}}}
\newcommand{\wup}{\left(1 + \frac{\tau^2}{\omega^2}\right)^{\frac{1}{2}}}
\newcommand{\ta}{\tilde{a}}
\newcommand{\tJ}{\tilde{J}}
\newcommand{\DI}[1]{\mbox{$\displaystyle{#1}$}}
\newcommand{\HHs}{\mbox{I\hspace{-.4ex}H }}
\newcommand{\HH}{\mbox{I\hspace{-.4ex}H}}
\newcommand{\wi}[1]{\widehat{#1}}
\newcommand{\SL}{\mbox{SL(2,\rm I\hspace{-.4ex}R)}}
\newcommand{\SLs}{\mbox{SL(2,\rm I\hspace{-.4ex}R) }}
\newcommand{\slalgs}{\mbox{sl(2,\rm I\hspace{-.4ex}R) }}
\newcommand{\slalg}{\mbox{sl(2,\rm I\hspace{-.4ex}R)}}
\newcommand{\scoset}{\mbox{SL(2,\rm I\hspace{-.4ex}R)/SO(2)}}
\newcommand{\scosets}{\mbox{SL(2,\rm I\hspace{-.4ex}R)/SO(2) }}


\thispagestyle{empty}

\hbox to \hsize{%
  \vtop{} \hfill
  \vtop{\hbox{MPI-PhT/96-53}\hbox{}}}

\vspace*{1cm}

\bigskip\bigskip\begin{center}
{\bf \Huge{ A Harmonic Space Approach to  Spherically 
Symmetric Quantum Gravity}}
\end{center}  \vskip 1.0truecm
\centerline{{\bf Helia Hollmann\footnote{e-mail: 
hollmann@mppmu.mpg.de}}}
\vskip5mm
\centerline{Max-Planck-Institut f\"{u}r
 Physik, Werner-Heisenberg-Institut}
\centerline{F\"ohringer Ring 6, 80805 Munich, Germany}
\vskip 2cm
\bigskip \nopagebreak \begin{abstract}
\noindent
After dimensional reduction the stationary spherically symmetric 
sector of Einstein's gravity is identified with an \scosets Sigma 
model coupled to a one dimensional gravitational remnant. 
The space of classical solutions consists of a one parameter 
family interpolating  between the Schwarzschild and the Taub-NUT 
solution. A Dirac Quantization of this system is performed and 
the observables -- the Schwarzschild mass and the Taub-NUT charge 
operator -- are shown to be self-adjoint operators with a 
continuous spectrum ranging from $-\infty$ to $\infty$. The 
Hilbert space is constructed explicitely using a harmonic space 
approach.

\end{abstract}

\newpage\setcounter{page}1

\section{Introduction}

The quantization of Einstein's gravity is one of the outstanding 
and most challenging areas in theoretical physics. There have 
been many attempts at a solution of this problem classified 
by Isham in his articles about quantum gravity \cite{Ish93}, 
\cite{Ish95}. One can start with the classical theory of general 
relativity and apply some kind of quantization algorithm. General 
relativity can be considered as a low-energy limit of another 
theory like superstring theory for example, one can try to force some 
principal ideas of a quantum field theory to be compatible with 
general relativity or the fundament of quantum gravity is seen to 
be some radically new perspective which leads in certain limiting 
situations to the classical theory of general relativity. Attempts 
in this direction are quantum groups or the noncommutative geometry 
invented by Connes. Nonperturbative approaches to the problem 
became of great importance since it is known 
that quantum gravity is non-renormalizable.  

This paper originates from a conservative point of view. Starting from 
the classical theory of general relativity one achieves a 
quantization via a canonical scheme, the ADM quantization. 
For full general relativity this is still an unsolved 
problem, but freezing all but a few of the infinitely many degrees of 
freedom of the gravitational field solutions for special sectors of 
Einstein's vacum theory are within reach. It is hoped, that at least 
some relevant features of the quantum theory are preserved. 
Even  restricting oneself to a particularly symmetric case, that is 
spherical symmetry -- as it is done here -- the space of classical 
solutions contains a black hole, the Schwarzschild solution, whose 
special importance is explained now. There are no particles in the 
sense of ordinary field theory. Even the concept of solitons 
has to be modified as shown in \cite{BreMaiGib88}, \cite{Car73}. 
One has to allow for 
singularities in the solutions leading to the suggestion that 
black holes are the proper particles of general relativity: 
coupling a point particle to its own gravitational field  
turns it into a black hole \cite{Dam86}. Furthermore classical 
black holes are stable under local perturbations 
\cite{Cha83}, and they have remarkable uniqueness properties. 
This is why they are considered to be the fundamental pieces of a 
quantum theory of gravitation and the investigation of the 
spherically symmetric sector of Einstein's theory of gravitation 
is seen in different a light \cite{HawEll73}, \cite{Car73}, 
\cite{BreMaiGib88}. As in conventional gauge theory, where 
one tries to keep manifest Poincar\'e invariance even in the process 
of gauge fixing, this sector of Einstein's gravity has a global 
symmetry group, the \SL, which will be preserved even  
at a quantum level. The unitary irreducible representations 
play a key role in the process of quantization. 
The major advantage of this model is that the 
quantum theory is explicitely known. 

One of the most important developments in field theory during the last 
two decades was the discovery of the quantum mechanical instability 
of black holes due to Hawking radiation. Most of the calculations 
of black hole radiation involve matter fields quantized in front of 
 a classical background. After the coupling of matter fields this 
model opens the possibility to take a closer look at 
Hawking radiation quantum-mechanically. But this is considered to 
be a project of future research.   
  
This paper mainly splits into a classical and a quantum part. 
In the beginning the spherically symmetric sector of Einstein's 
gravity is introduced and a Kaluza-Klein like reduction is 
performed \cite{BreMai86}. This treatment is rather concise, but 
a more comprehensive and more general exposition can be found in 
\cite{BreMai86}. It turns out, that gravity reduced from four 
to three dimensions in its dual representation is equivalent to a 
\scosets nonlinear coset sigma model coupled to three dimensional 
gravity. Using this coset space structure all stationary spherically 
symmetric solutions of Einstein's equations with spherical symmetry 
are constructed in \cite{DobMai82}: there are the Schwarzschild 
and the Taub-NUT solution and a one-parameter family connecting 
these two \cite{Tau51}. The parameters of the theory are hidden 
in the sigma model currents: the Schwarzschild mass $m$ and the 
Taub-NUT charge $l$. In the following the problem is treated as a 
constraint system. The Hamiltonian is the only constraint which  
survives the process of dimensional reduction. The components of 
the sigma model currents consequently turn out to be observables 
in the sense of Dirac and form an \slalgs algebra. 

During quantization the classical phase space is turned 
into a Hilbert space on which the operators - the former functions 
on phase space - act. In general it is a difficult problem to 
identify the Hilbert space. On the other hand it is the Hilbert space,     
which is of fundamental importance to make assertions about the operators 
to be considered physical. The implementation of the classical \SLs  
symmetry at a quantum level leads to a solution of this problem. 
The eigenfunctions of the invariant differential operator on the 
group span the Hilbert space as is shown by the application of a 
generalized Plancherel theorem. 
The measure on the Hilbert space is derived quite naturally from the 
Haar measure on the group. 
Once knowing the Hilbert space explicitly 
one can investigate the self-adjointness and the spectrum of certain 
differential operators. Both the mass operator and the charge 
operator are self-adjoint and their spectra are purely continuous. 
Neither the spectrum of the mass operator nor the spectrum of 
the charge operator are bounded. That is: negative Taub-NUT charges 
as well as negative masses belong to the spectrum, too. 

The most challenging feature of this model is that bases of 
the Hilbert space are at hand. Therefore the model serves as a  
kind of test laboratory for several aspects of quantum gravity. 
But these topics will be reserved for a forthcoming publication.     

This method is not limited to 
the situation here. It is applicable to various other models 
occurring in conformal field theory, string theory, quantum 
cosmology or supergravity. 
\section{The Classical Theory}

This classical part starts with the application of a Kaluza-Klein 
reduction \cite{Kal21} to a four dimensional Einstein space 
$M_4$ with signature $(+, +, +, -)$  having a one parameter 
Abelian isometry group \cite{DobMai82} which acts freely and  
corresponds to time translations. That is there exists a timelike 
Killing vector field $K^{^M}$ describing the action of the Lie 
algebra of the isometry group. In a coordinate basis 
the metric $g_{_{MN}}$ of $M_4$ decomposes into a metric 
$h_{mn}$ on the remaining three spatial dimensions, a scalar field 
$\tau$ and the vector field $B_m$:       
\[
  g_{_{MN}} =
  \left(
    \btab{c|c}
      $\DI{-\frac{1}{\tau}} h_{mn} 
      + \tau B_m B_n$ & $-\tau B_n$ \\[1.5ex] \hline 
      $-\tau B_m$  & $\tau$
    \etab
  \right).
\]
The extra factor $\tau$ assures that diffeomorphisms of the 
special form $t \ms t + \gL(x^m)$ act as gauge transformations 
$B_m \ms B_m + \pa_m \gL$  on the vector field $B_m$ 
\cite{BreMaiGib88}. Capital letters $M, N$ vary from one to 
four and small ones from one to three. Plugging this Ansatz 
for the four dimensional metric into the Einstein Hilbert action   
\be 
  S = - \frac{1}{2} \: \int d^4 x \: \sqrt{-g} \: R 
\ee
leads to the Lagrangian 
\be
  \label{l43}
  {\cal L}^{(4,3)} = - \frac{1}{2} \sqrt{-g} \: R = \sqrt{h} 
  \left(
    - \frac{1}{2} {}^{(3)} R - \frac{\tau^2}{8} F^{mn} F_{mn} 
    - \frac{1}{4 \tau^2} \pa^m \tau \pa_m \tau 
  \right).
\ee 
$F_{mn}$ denotes the field strenght of $B_m$, defined by 
$F_{mn} = \pa_m B_n - \pa_n B_m$. $R,\: {}^{(3)}R$ are the scalar    
curvatures corresponding to the four dimensional metric $g_{_{MN}}$ 
and the three dimensional metric $h_{mn}$, respectively. 
$\sqrt{-g}$ and $\sqrt{h}$ are the square roots of the 
determinants of the metrics as usual. 

Under the specified conditions dimensional reduction of four 
dimensional pure gravity to three dimensions leads to gravity 
coupled to an abelian vector field and a scalar.   

The equations of motion derived from (\ref{l43}) can be interpreted 
as integrability condition for the vector field $B_m$: 
\[ 
  \tau^2 F_{mn} = \epsilon_{mnp}~\pa^p \go. 
\]
Elimination of the field 
strength in terms of the gravito-magnetic potential $\go$ leads 
to a dual Lagrangian density 
\[ 
  {\cal L}^{^D} = \sqrt{h} 
  \left(
    - \frac{1}{2} {}^{(3)}R ~+~ \frac{h^{mn}}{4 \tau^2} 
    \left(
      \pa_m \tau \pa_n \tau + \pa_m \go \pa_n \go
    \right)
  \right),
\]
which can be rewritten as 
\be
  \label{ld43}
  {\cal L}^{^D} = \sqrt{h} 
  \left(
    - \frac{1}{2} {}^{(3)}R ~+~ \frac{h^{mn}}{8} \mbox{\rm Tr}
    \left(
      \chi^{-1} \pa_m \chi \: \chi^{-1} \pa_n \chi
    \right)
  \right),
\ee
where
\[
  \chi = 
  \left(
    \ba{cc}
      \tau + \DI{\frac{\go^2}{\tau}} &  \DI{\frac{\go}{\tau}} \\
      & \\ 
      \DI{\frac{\go}{\tau}}  & \DI{\frac{1}{\tau}}  \\
    \ea
  \right).
\]
That is four dimensional reduced 
gravity in its dual representation is equivalent to an  
\scosets sigma model coupled to three dimensional 
gravity. The matrix $\chi$ is an element of the Riemannian 
symmetric space \scoset. The fields contained in $\chi$ have a 
physical interpretation: The norm $\tau$ of the 
Killing vector plays the role of a gravitational potential,  
and $\go$ is the so called gravito-magnetic or NUT potential. 

For the derivation of the equations of motion one starts from  
(\ref{ld43}). 
$R_{_{MN}} = 0$ is equivalent to the set of equations 
\bsea
  \label{eom1}
  && ^{(3)} R_{mn} = \frac{1}{2} \mbox{\rm Tr} 
  \left(
     \chi^{-1} \pa_m \chi \: \chi^{-1} \pa_n \chi  
  \right) \\
  \label{eom2} 
  && D^m \left( \chi^{-1} \pa_m \chi \right) = 0
\esea

Imposing an additional SO(3) symmetry on the remaining 
three spatial dimensions means to constrain the involved fields 
in such a way that they depend on one spatial dimension only:
$f, \tau$ and $\go$ are functions of $\rho$. In this stationary 
case the NUT potential can be considered as a kind of ``magnetic 
potential'' in analogy to Maxwell's theory. 

It is convenient to parametrize the metric $h_{mn}$ by polar 
coordinates
\be
\label{g3}  
  h_{mn} = 
  \left(
    \ba{ccc}
      N^2(\rho) & & 0 \\
      & f^2(\rho) & \\
      0 & & f^2(\rho) \sin^2\gt 
    \ea
  \right) , \qquad m,n = \rho, \gt, \gp
\ee
After substitution of this metric into the Lagrangian one obtains 
\be
  \label{lm}
  {\cal L}^{^D} = N \left[ 
  \frac{f^{\prime 2}}{N^2} + 1 - \frac{f^2}{4 N^2 \tau^2} 
  \left(
    \tau^{\prime 2} + \go^{\prime 2} 
  \right) \right].
\ee
The prime $^\prime$ denotes the derivative with respect to $\rho$.
$N$ is the ``lapse'' function. ``lapse'' is set in quotation marks, 
because it usually refers to a timelike direction whereas the 
lapse indicates spacelike propagation.  

For stationary spherically symmetric gravity 
$\chi^{-1} \chi^\prime$ is unequal to zero and (\ref{eom1}) 
simplifies to 
\bea
  ^{(3)}R_{22} &=& ^{(3)}R_{33} = 
  \left[
    \frac{f^{\prime \prime}}{f} + 
    \left( \frac{f^\prime}{f} \right)^2 
    - \frac{1}{f^2} 
  \right] = 0 \\ \nonumber
  ^{(3)}R_{11} &=& - \frac{2 f^{\prime \prime}}{f} 
  = \frac{1}{4} \mbox{Tr} ( \chi^{-1} \chi^\prime)^2 \\ \nonumber 
  (f^2 \chi^{-1} \chi^\prime )^\prime &=& 0. \nonumber
\eea  
The lapse function refers to a gauge degree of freedom and was set 
equal to one meanwhile.  
Dobiasch and Maison \cite{DobMai82} calculated the solution of these 
equations of motion. They found 
\be
  \label{st1}
  f^2(\rho) = R^2 - a^2, \qquad R = \rho - b. 
\ee
$a, b $ are constants of integration and will be interpreted later. 
It turns out that $\chi$ is of the form 
\[ 
  \chi = \chi_0 ~ \me^{t(\rho)\:\mu}, \quad \mbox{with} \quad \chi_0 = 
  \left(
    \ba{cc}
      1&0\\
      0&1
    \ea
  \right) \quad \mbox{and} \quad 
  t(\rho) = - \int_\rho^\infty f^{-2}(s) ds. 
\]
The most general matrix $\mu$ can be written as 
\[ 
  \mu = \sin\psi ~ \mu_1 + \cos\psi ~ \mu_2, \quad \mbox{with} \quad 
  \mu_1 = 
  \left(
    \ba{cc}
      0 & a \\
      a & 0 
    \ea
  \right) \qquad \mbox{and} \qquad \mu_2 = 
  \left(
    \ba{cc}
       - a & 0 \\
         0 & a 
    \ea
  \right).
\]
Using this information the classical fields $\go$ and $\tau$ 
are derived:
\bsea 
  \label{st2}
  \tau &=& \frac{R^2 - a^2}{R^2 + a^2 + 2 \cos \psi R a} \\
  \go &=& \frac{2 \sin\psi R a}{R^2 + a^2 + 2 \cos \psi R a}
\esea
For the sake of completeness we also give the four dimensional 
metric in terms of the one dimensional data:
\[
  g_{_{MN}} = 
  \left(
    \ba{cc}
      \ba{cc}
        - \DI{\frac{1}{\tau}} & \\
        & - \DI{\frac{f^2}{\tau}}
      \ea & 0 \\ 0 &
      \ba{cc}
        - \DI{\frac{f^2}{\tau}} ~\sin^2\gt 
        + \DI{\frac{f^4}{\tau^3}} ~ \cos^2 \gt ~\go^{\prime 2} & 
        \DI{\frac{f^2}{\tau}} ~ \cos\gt ~ \go^\prime\\
        \DI{\frac{f^2}{\tau}} ~ \cos \gt ~ \go^\prime & \tau 
      \ea
    \ea
  \right), 
\]  
$M,N = \rho, \gt, \gp, t$. 
The solution corresponding to $\sin\psi = 0$ turns out to be the 
Schwarzschild solution, which is the gravitational field outside 
a spherically symmetric mass distribution. It is a static 
solution of Einstein's equations. For $\cos\psi = 0$ only the 
$\mu_1$ part remains and it is a Taub-NUT solution. The geometrical 
difference between the Schwarzschild and the Taub-NUT solution is, 
that the latter admits an  
isometry group whose orbits are on 3-spheres, whereas in 
the case of the Schwarzschild solution they are 2-spheres. The 
``result'' is the Taub-NUT time being a circle. Another 
interpretation is given in terms of the difference between staticity 
and stationarity: In the static case there are two symmetries: 
a time translation symmetry and a time reflection symmetry. For 
Taub-NUT the fields being time translation invariant fail to be 
time reflection invariant: the neighbouring orbits of the Killing 
vector fields twist around each other!
Both solutions are asymptotically flat, but only the Schwarzschild 
solution is also asymptotically Minkowski.

Although it should be clear from the structure of the Lagrangian 
(\ref{ld43}) it is stressed at this point once more, that the space 
of solutions contain more than the Schwarzschild solution. 
The attempts to quantize spherically symmetric gravity 
\cite{CavdeAFil951}, \cite{CavdeAFil952}, \cite{KasThi93}, 
\cite{Kuc94}, \cite{MarGegKun94} essentially refer to the static 
truncation GL(1) of this \scosets sigma model.    

In order to stress the one to one correspondence between the 
observables and the initial data it is shown (see \cite{DobMai82}) 
how to use the coset space structure to obtain (\ref{st2})

The invariant line element of the coset is
\be
  \label{come}
  ds^2 = \frac{1}{4} \mbox{Tr}(\chi^{-1} d \chi )^2
   = \frac{1}{2 \tau^2} (d\tau^2 + d\go^2).
\ee
The geodesic motion is defined by
\[
  \frac{d^2 \Phi^i}{d\gs^2}
   + \gC^i{ }_{jk} \frac{d \Phi^j}{d\gs} \frac{d \Phi^k}{d\gs} = 0.
\]
Here $\Phi^i$ are the coordinates $\tau$ and $\go$, $\gC^i{ }_{jk}$
are the Christoffel symbols with respect to the metric defined
by the line element (\ref{come}) above.
This leads to the following system of differential equations
\bea
  \label{ge}
  \ddot{\tau} &-& \frac{\dot{\tau}^2}{\tau} 
               + \frac{\dot{\go}^2}{\tau} = 0, \\
  \ddot{\go} &-& \frac{2 \dot{\tau} \dot{\go}}{\tau} = 0. \nonumber
\eea
The dot $\dot{}$ denotes differentiation with respect to $\gs$.
In order to guarantee that $ \lim_{\rho \ra \infty} \chi_0$ is
equal to the identity, one chooses the boundary values
$\tau(\gs=0) = 1 $ for the gravitational
potential and $\go(\gs=0) = 0$ in the case of the gravito-magnetic 
potential. To relate the geodesic variable somewhat more directly 
to the original
$\rho$-variable, we cite the solution for $a^2 > 0$
and $a = 0$ from \cite{DobMai82}. If $a^2 > 0$ the function 
$f^2(\rho)$ has two simple zeros at $\rho = b \pm a$ causing
\[
  \gs(\rho) = \frac{1}{a} \ln
  \left(
    \frac{\rho - b - a}{\rho - b + a}
  \right), 
\]
which tends to $- \infty$ at $\rho = b - a$. Therefore $\gs=0$ 
corresponds to infinite value of $\rho$ with the interpretation that at
$\gs = 0$ a geodesic starts at $(\tau_0, \go_0)$ and approaches some
other value $(\tau_h, \go_h)$ at $\gs = \infty$ which marks
the position of the horizon of a black hole.
$a = 0$ means Schwarzschild mass $m$ and Taub-NUT charge $l$ equal to
zero and for this reason represents flat space.

The system of differential equations (\ref{ge}) is easy to solve:
\bea
  \go(\gs) &=&  \cosh \gb \: \tanh(a \gs + \gb) \:-\: \sinh \gb, \\
  \tau(\gs) &=& \frac{\cosh \gb}{\cosh(a \gs + \gb)}.
\eea
The boundary conditions have to be taken into account.

To summarize: each geodesic corresponds to a solution with a
required asymptotic behaviour, namely $(\tau_0, \go_0) =  (1,0)$.
It is uniquely determined by its ``velocity''
\[
  (\dot{\tau}_0, \dot{\go}_0)
  \:=\: (- a \tanh \gb, \frac{a}{\cosh} \gb) = (m,l).
\]

After this excursion the Lagrangian (\ref{ld43}) or (\ref{lm}), 
respectively is investigated once more. It is invariant under \SLs 
transformations $z \ms \frac{az + b}{cz + d},\: z = \go + i \tau, \: 
ad - bc = 1$. The corresponding Noether currents $J$ are  
\[
  J = 
  \left(
   \ba{ll}
     -J^0 & J^+ \\
     J^- & J^0 
   \ea
  \right) := \frac{f^2}{2 N^2} \chi^{-1} \chi^\prime  = 
  \left(
    \ba{cc}
      \DI{- \frac{\tau^\prime}{\tau} - \frac{\go \go^\prime}{\tau^2}} & 
      \DI{\frac{\go^\prime}{\tau^2}} \\
      & \\
      \DI{\go^\prime - \frac{2 \go \tau^\prime}{\tau} 
      - \frac{\go^2 \go^\prime}{\tau^2}} & 
      \DI{\frac{\go \go^\prime}{\tau^2} + \frac{\tau^\prime}{\tau}} 
    \ea
  \right)
\]
As $\chi$ is a symmetric matrix only two components of the current 
matrix are linearly independent. 
Imposing proper boundary conditions at infinity on the sigma model 
fields there exists an asymptotic multipole expansion of $\chi$. 
\[
  \chi \sim \sum_{n=0}^{\infty} \rho^{-n} ~ \chi_n(\gt, \gp).
\]
A suitable choice of coordinates asymptotically leads to 
\[ 
  \go = \frac{l}{\rho} + O(\frac{1}{\rho^2}) 
  \qquad \mbox{and} \qquad  
  \tau = 1 - \frac{2m}{\rho} + O(\frac{1}{\rho^2}).
\]
One is interested in particular in the $1/\rho$ term in the expansion of 
$\chi$ because it contains the parameters, the Schwarzschild mass 
$m$ and the Taub-NUT charge $l$, which are the entries of  
the matrix of global charges  
\[
  Q = \frac{1}{4 \pi} \int_{\pa S^2} J_\rho ~ d\Sigma^\rho 
    = \chi_0^{-1} \chi_1. 
\]
By integration over an infinitely large sphere $Q$ is calculated to be 
\[
  Q =  
  \left(
    \ba{cc}
      m & l \\
      l & -m 
    \ea
  \right).
\]
The identification of the Schwarzschild mass $m$ and the Taub-NUT 
charge $l$ is obtained by comparison with the standard form of 
the Schwarzschild and the Taub-NUT solution in the literature 
\cite{Kraetal80}.   
The $J^0$ component of the current is related to the 
Schwarzschild mass $m$, and the $J^+$ and $J^-$ components lead to 
the Taub-NUT charge $l$. 

The line element $ \frac{f^4}{4 N^4} \mbox{Tr}(\chi^{-1} d\chi )^2$ 
is constant on geodesics and assumes values greater than or equal 
to zero. From an algebraic point of view $\mbox{Tr}J^2$ is the quadratic 
Casimir element. Here it leads to a generalized charge conservation law: 
As explained above the mass $m$ and the Taub-NUT parameter $l$ are 
viewed as generalized charges. 
On the space of classical solutions the invariant line element 
is evaluated to be $4 a^2 = m^2 + l^2$.

So far the space of classical solutions is understood completely. 
Now one turns to the ADM approach to Einstein's gravity. 
General relativity is invariant with respect to space-time 
diffeomorphisms. This local symmetry relates part of the solutions 
stemming from the same initial conditions. In the Lagrangian 
formalism this results in the fact that out of ten field equations 
there are only six independent ones: The reparametrization 
invariance leads to Noether currents, the so called Bianchi 
identities, and in the sequel to the presence of an arbitrary 
function of time in the general solution of the equations of 
motion. A solution is mapped to a solution by a gauge 
transformation. Performing a Hamiltonian formalism   
the local symmetry transformations yield a system with 
constraints. This means that there are conditions on the 
allowed initial data which must be preserved during 
time evolution. This consistency requirement can lead to 
secondary and higher order constraints. The gauge degrees 
of freedom are hidden in the system of first class constraints, 
these are those whose  Poisson bracket vanish weakly, 
that is on-shell. 

Here a modified Hamiltonian formalism is implemented. 
The slicing is performed according to the $\rho$ - that is a 
spacelike - coordinate. How to proceed in this case can be found 
in \cite{Mar94}, \cite{CavdeAFil951}. Due to spherical symmetry 
the lapse function is the only Lagrange multiplier which survives. 
In other words: 
it expresses the invariance under $\rho$-reparametrization and this 
leads to the only primary first class constraint of the 
theory, the Hamiltonian constraint, which generates the gauge 
transformations    
\[
  H = \frac{1}{4} \pi_f^2 - 1 
      - \frac{\tau^2}{f^2} ( \pi_\tau^2 + \pi_\go^2).
\]
In terms of the fields and their conjugate momenta $J$ reads
\be
  \label{ic}
  J = 
  \left(
    \ba{ll}
      - \tau \pi_\tau - \go \pi_\go & \pi_\go \\
      \tau^2 \pi_\go - 2 \go \tau \pi_\tau - \go^2 \pi_\go & 
      \tau \pi_\tau + \go \pi_\go 
    \ea
  \right).
\ee
By definition an observable is a function on the constraint surface 
that is gauge invariant. As explained above it has weakly, i.e. 
on-shell, vanishing brackets with the first class constraints.  
Observables do not evolve in ``time'' and therefore there is a 
one to one correspondence between the observables and the initial 
data here: They can be identified with the 
space of solutions. Here the classical solutions are parametrized 
by the Schwarzschild mass $m$ and the Taub-NUT charge $l$. Therefore 
one expects the existence of two observables, which turn out to be 
``hidden'' in the current matrix $J$:  
\[
  \{ H, J^0 \} = 0, \qquad \{ H, J^+ \} = 0, 
  \qquad \{ H , J^-\} = 0. 
\]
On the other hand they generate an \slalgs 
algebra:
\[
  \{ J^+, J^0 \} = -J^+, \qquad \{ J^+, J^- \} = 2 J^0, 
  \qquad \{ J^0 , J^-\} = -J^-. 
\]
The observable $J^0$ measures the Schwarzschild mass, 
$J^+$ and $J^-$ lead to the Taub-NUT charge $l$ and $\mbox{Tr}J^2$ 
yields the value of the invariant line element. 
Because $\mbox{Tr}J^2$ commutes with all the currents 
\[
  \{ H, \mbox{Tr}J^2 \} = 0, \quad \{ \mbox{Tr}J^2, J^+ \} = 0, 
  \quad \{ \mbox{Tr}J^2 , J^-\} = 0, \quad  
  \quad \{ \mbox{Tr}J^2 , J^0\} = 0,   
\]
these observables can be ``measured simultaneously'' 
(simultaneously diagonalized).
A ``simultaneous measurement'' of the Schwarzschild mass $m$ and the 
Taub-NUT charge $l$ is not possible.

The Hamiltonian stated above is of course not the whole story 
\cite{RegTei74}, \cite{BeiMur87}. To develop a Hamiltonian   
formulation that is Poincar\'e invariant at infinity one needs 
a more precise specification of the asymptotic form of the 
canonical variables \cite{RegTei74}, \cite{Kuc94}. Even more 
is true: in general the usual Hamiltonian in the case of an 
open universe does not have well defined functional derivatives 
and consequently 
such a Hamiltonian does not generate any equation of motion 
at all. To develop a Hamiltonian theory which is Poincar\'e 
invariant at infinity one has to subtract some boundary terms 
arising as ten new constraints with arbitrary multipliers 
in the Hamiltonian. Physically these constraints are related 
to the energy, the total momentum and the angular momentum 
of the space-time under consideration. The surface integral 
leads to a specific asymptotic behaviour of the fields. 
But - as outlined by Regge and Teitelboim \cite{RegTei74} - 
if one assures that the lapse and the shift functions have a   
proper asymptotic behaviour, one can consider the Hamiltonian 
and the Diffeomorphism constraints (the later vanish in this 
sector of Einstein's gravity) without their asymptotic ends. 
This is what is done here.    

\section{Quantization}

Concerning the canonical approach to quantum gravity there are two 
main streams to handle the problem \cite{HenTei92}: 
the reduced phase space 
quantization and the Dirac quantization. From a ``Hamiltonian 
point of view'' the former deals with the problem by elimination 
of the first class constraints at an early stage. This amounts to  
quantize gauge invariant functions only, i.e. constants of motion. 
To carry out the reduced phase space quantization one must find a 
complete set of gauge invariant functions, which is a difficult 
task in general. 
Therefore often canonical gauge conditions are imposed to obtain  
the reduced phase space. After completely fixing the gauge any function of 
the canonical variables can be viewed as the restriction of a gauge 
invariant function in that gauge. Hence: once the gauge is fixed 
one is effectively working with gauge invariant functions. A complete 
set of independent gauge fixed functions provides a complete 
set of gauge invariant functions. This method is practicable if there 
are no Gribov obstructions. 

Problems of that quantization scheme are that an early elimination 
of the gauge degrees of freedom may spoil the manifest invariance 
under important symmetries and in general it destroys locality in 
space. Furthermore it might happen that the brackets of the 
observables are complicated functions and the question arises how 
to realize them quantum mechanically. 
The main advantage is that only the physical degrees of freedom 
are quantized. Every state in the Hilbert space is a physical 
one. 

In the so called Dirac quantization one keeps the gauge degrees of 
freedom. The classical phase space functions become operators acting 
in the Dirac representation space, which carries  
nonphysical information. Hence one has to select a physical 
subspace of gauge invariant states. 

This paper proceeds along the Dirac approach. The \SLs symmetry of 
the theory provides the key to quantize this sector of Einstein's gravity. 
An outline of what happens next is given: The former phase space functions 
become operators. After this step a naiv approach is followed: 
the Wheeler-DeWitt is solved formally. For the investigation of the 
self-adjointness and the spectrum of the physical operators it is 
necessary to specify the Hilbert space by employing group theoretical 
arguments. 
Then, to justify the choice of the Hilbert space or to obtain a basis  
harmonic analysis is heavily used. That is what is done finally in 
this quantum part of the paper.   

During the process of quantization the functions $J^+, J^-, J^0, 
\mbox{Tr}J^2$ and $H$ on phase space become operators 
on an appropriate Hilbert space. 
In the Schr\"{o}dinger representation the
fields are turned into multiplication operators 
and the momenta become differentiation operators
$\pi_\Phi \ms -i \pa_\Phi$, where $\pa_\Phi$ is an abbreviation
for $\frac{\pa}{\pa_\Phi}$.  
The generators of the currents $J^0, \: J^+, \: J^-,$  
the Casimir Tr$J^2$ and the Hamiltonian $H$ become: 
\bes
  \wi{J^0} &=& -i \: \tau \pa_\tau \:-\: i \: \go \pa_\go \\
  \wi{J^+} &=& i \: \pa_\go \\
  \wi{J^-} &=& i \: (\tau^2 \:-\: \go^2) \: \pa_\go
     \:-\: 2 i \: \go \tau \: \pa_\tau \\
  \la &=& - \: \tau^2 \:(\pa_\tau^2 \:+\: \pa_\go^2) \\
  \wi{H} &=& - \frac{\pa_f^2}{4} \:+\: \frac{1}{2 f^2} \:
     \la \:-\: 1.
\ees
Applied to a wave function $\psi$ the Hamiltonian defines the so
called Wheeler-DeWitt equation $\wi{H} \psi = 0$.
The following commutation relations hold:
\[
  [\: \wi{H}, \: \wi{J^0}] \:=\: 0 \qquad
  [\: \wi{H}, \: \wi{J^+}] \:=\: 0 \qquad
  [\: \wi{H}, \: \wi{J^-}] \:=\: 0,
\]
i.e. the current operators are observables. The Casimir operator
$\la$ commutes with the Hamiltonian and the currents
\[
  [\: \la, \: \wi{H}] \:=\: 0 \qquad
  [\: \la, \: \wi{J^0}] \:=\: 0 \qquad
  [\: \la, \: \wi{J^+}] \:=\: 0 \qquad
  [\: \la, \: \wi{J^-}] \:=\: 0,
\]
and $\wi{J^0}, \wi{J^+}, \wi{J^-}$ form an \slalgs algebra:
\[
  [\: \wi{J^+}, \: \wi{J^0}] \:=\: - i \wi{J^+}, \qquad
  [\: \wi{J^+}, \: \wi{J^-}] \:=\: 2 i \: \wi{J^0}, \qquad
  [\: \wi{J^0}, \: \wi{J^-}] \:=\: - i \wi{J^-}.
\]
The Casimir operator $\la$ is an essential part of the Wheeler-DeWitt
equation. As it commutes with the current operators, $\la$ and any one
of the $\wi{J}$'s can be ``measured simultaneously'', i.e. they can be
simultaneously diagonalized. $\wi{J^0}, \wi{J^+}$ and $\wi{J^-}$ do not
commute with each other. Therefore even formally a direct
``measurement'' of the Schwarzschild mass $m$ and the Taub-NUT 
charge $l$ is not possible.

The interpretation of the current operators as observables
of the theory strongly suggests to preserve the \SLs symmetry
at the quantum level: $\wi{J^0}, \wi{J^-}, \wi{J^+}$ and the 
Casimir operator $\la$ are forced to become self-adjoint operators 
during the process of quantization. That is, the quantum mechanical Hilbert
space is built from the unitary irreducible representations of the group
\SL. Finding a solution of the Wheeler-DeWitt equation therefore 
basically becomes a group theoretical problem as the equation  splits into
an $f$ dependent and an $f$ independent part, which is the Casimir
operator on the group up to a constant term.

There are two particularly useful possibilities to diagonalize
the Casimir operator $\la$.
On one hand one solves the differential equations for $\la$ 
and the Taub-NUT charge operator $\wi{J^+}$ simultaneously. 
The other possibility is to diagonalize the Schwarzschild mass
operator $\wi{J^0}$ and the Casimir operator $\la$. 

The former case is investigated first, i.e. the formal solutions of
the following system of differential equations
\bsea
  \label{dgl1}
  - \tau^2 \left( \pa_\tau^2  \:+\: \pa_\go^2 \right) \:
    \psi_{\gl L}(\go, \tau)
  & = & \gl \: \psi_{\gl L}(\go, \tau), \\
  i \: \pa_\go \: \psi_{\gl L}(\go, \tau)
  & = & L \: \psi_{\gl L}(\go, \tau).
\esea
are obtained. The second equation (\ref{dgl1}b) is considered first.
The solution is
\[
  \psi_{\gl L}(\go, \tau) \:=\: C(\tau) \: \me^{-i L \go}.
\]
Substitution of this solution into the first equation yields
\[
  \tau^2 \: \pa_\tau^2 \: C(\tau) \:+\:
  \left( \gl \:-\: \tau^2 L^2 \right) \: C(\tau) \:=\: 0.
\]
This is a differential equation of Bessel type.
If $L^2 \neq 0$, it follows that
\be
  \label{bes1}
  C(\tau) \:=\: \sqrt{\tau} \:
  \left(
    \hat{C}_1 \: \mbox{J}_k(i |L| \: \tau) \:+\:
    \hat{C}_2 \: \mbox{Y}_k(i |L| \: \tau)
  \right), \qquad k \:=\: \frac{1}{2} \sqrt{1 - 4 \gl}
\ee
where J$_k$ denote the Bessel functions of the first kind and
Y$_k$ those of the second kind.
Later on it will be convenient to use a different linear combination 
of the fundamental solutions. Therefore the following relations 
between different types of Bessel functions are used: 
\bes
  \mbox{J}_k( i z) &=& \me^{^\frac{i k \pi}{2}} \: \mbox{I}_k(z) \\
  \mbox{Y}_k(i z) &=& \me^{^\frac{i (k+1) \pi}{2}} \: \mbox{I}_k(z)
    \:+\: \frac{2}{\pi} \: \me^{^{-\frac{i k \pi}{2}}} \: \mbox{K}_k(z)
\ees
With suitably chosen constants $C_1$ and $C_2$
the original formal solution (\ref{bes1}) can be written as:
\[
  \psi_{\gl L}(\go, \tau) \:=\: \psi_{\gl L}^1(\go, \tau)
      \:+\: \psi_{\gl L}^2(\go, \tau)
\]
with
\bsea
\label{bes2}
  \psi_{\gl L}^1(\go, \tau) &=&
   C_1 \: \me^{-i L \go} \: \sqrt{\tau} \: \mbox{I}_k(|L| \: \tau), \\
  \psi_{\gl L}^2(\go, \tau) &=&
   C_2 \: \me^{-i L \go} \: \sqrt{\tau} \: \mbox{K}_k(|L| \: \tau).
\esea
I$_k$ and K$_k$ denote the Bessel functions of imaginary argument.
%

Now the simultaneous diagonalization of the Casimir operator
$\la$ and the mass operator $\wi{J^0}$ is carried out.
As in the former case the system of differential equations
\bsea
  \label{dgl2}
  - \tau^2 \: ( \pa_\tau^2 \:+\: \pa_\go^2 ) \:
  \psi_{\gl M}(\go, \tau)
  &=& \gl \: \psi_{\gl M}(\go, \tau) \\
  - i \: \left( \tau \pa_\tau \:+\: \go \pa_\go \right) \:
  \psi_{\gl M}(\go, \tau)
  &=& M \: \psi_{\gl M}(\go, \tau)
\esea
has to be solved. The solution of (\ref{dgl2}b) is
\be
\label{sinsol}
  \psi_{\gl M}(\go, \tau) \:=\:
  C\left(v \right) \: \go^{i M},
\ee
where $v = \frac{\tau}{\go}$. Alternatively $\frac{1}{v}$ 
could be used as independent variable.  
Nevertheless it turns out that (\ref{sinsol}) leads to a nice 
representation in terms of associated Legendre polynomials.
Substituting (\ref{sinsol}) into (\ref{dgl2}a) yields
\[
  - v^2 \: ( v^2 + 1 ) \: C^{\prime \prime}
  \:+\: 2 m v^3 \: C^\prime \:-\:
  \left(m ( m+1)\: v^2 \:- \: \gl \right) \: C \:=\: 0.
\]
Here the prime $^\prime$ denotes differentiation with respect 
to $v$ and $m$ is defined to be $ m := iM -1$. 
This differential equation is of Hypergeometric type and
transformed into the Hypergeometric differential equation
by
\[
  C(v) \:=\: v^{^\frac{1 + \gk}{2}} \: \eta(-v^2), \qquad
  \gk \:=\: \sqrt{1 - 4 \gl},
\]
where $\eta(\xi)$ is the solution of
\[
  \xi \: (\xi - 1) \: \eta^{\prime \prime} \:+\:
  [( \ga + \gb + 1) \: \xi \:-\: \gc ] \: \eta^\prime
  \:+\: \ga \gb \: \eta \:=\: 0
\]
The prime $^\prime$ denotes differentiation with respect to $\xi$,
and $\ga, \: \gb, \: \gc$ are equal to
\bes
  \ga &=& \frac{1}{4} \: ( 1 + \gk - 2m) \\
  \gb &=& - \frac{1}{4} \: ( 1 - \gk + 2m) \\
  \gc &=&  1 + \frac{\gk}{2}
\ees
If $ \gl \neq - n^2 + 1/4, \: n \in \Z$, $\eta(\xi)$
is calculated to be
\be
  \label{hyp2}
  \eta(\xi) \:=\: C_1 \: \mbox{F}(\ga, \gb; \gc; \xi)
  \:+\: C_2 \: \xi^{1 -\gc} \:
  \mbox{F}( \ga - \gc + 1, \gb - \gc + 1; 2 - \gc; \xi)
\ee
(\ref{hyp2}) can be expressed
by associated Legendre polynomials so that one finally
obtains
\[
  \psi_{\gl M}(\go, v) \:=\: \psi_{\gl M}^1(\go, v)
      \:+\: \psi_{\gl M}^2(\go, v)
\]
with
\bsea
\label{le1}
  \psi_{\gl M}^1(\go, v) \!\!\!\!\! &=&\!\!\!\!\!
     C_1 \: \go^{i M}  \sqrt{v}
     \left( 1 + v^2 \right)^{\frac{i M}{2} - \frac{1}{4}}
     \mbox{P}^{-\frac{k}{2}}_{-i M -\frac{1}{2}}
     \left( \frac{1}{\sqrt{1+v^2}} \right),  \\
  & & \nonumber \\
  \psi_{\gl M}^2(\go, \tau) \!\!\!\!\! &=&\!\!\!\!\!
     C_2 \: \go^{i M} v^{k + \frac{1}{2}} |v|^{-k}
     \left( 1 + v^2\right)^{\frac{i M }{2} - \frac{1}{4}}
     \mbox{P}^{\frac{k}{2}}_{-i M  - \frac{1}{2}}
     \left( \frac{1}{\sqrt{1 + v^2}} \right), 
\esea
where  $\gl \neq \frac{1}{4} - n^2, \quad n \:\in \: \Z.$

In the coordinates $\go$ and $v$ the differential operators
can be written in the following manner:
\bsea
  \label{dgl2b}
  \wi{J^0} &=& -i \go \: \pa_\go\\
   & & \nonumber \\
  \la &=& v^2 \: (1+v^2) \: \pa_v^2 \:+\: \go^2 v^2 \: \pa_\go^2
  \:-\: 2 v \go^3 \: \pa_\go \pa_v \:+\: 2 v^3 \: \pa_v
\esea
To solve (\ref{dgl2b}) means seeking solutions of
$\la \: \psi \:=\: \gl \psi$ of the form $ f(\go) C(v)$.

Of course, the formal solutions of the differential equations 
do not yield enough structural elements to solve the problem. 
For a deeper understanding a Hilbert space structure is 
needed to single out part of the solutions of the differential 
equations and to show self-adjointness of the physical operators. 
For this one applies various techniques from functional analysis, 
the representation theory of \SL, the Plancherel theorem, and its 
version on the coset space \scoset. 

The natural choice of the Hilbert space is to take the closure 
of the functions on the coset space \scosets on which the physical 
operators can be shown to be essentially self-adjoint. 
What one also would like to have is a subset of functions 
which approximate all functions on the coset space uniformly. 
This is the content of the theorem by Stone and Weierstrass. 
It states, that given a compact metric space $E$, which is 
constructed below, any subalgebra 
of continuous functions on that space which contains the unity 
and separates the points of $E$ is dense in the Banach space 
of functions $E$. One immediately recognizes that the space of 
$L^2$--functions over the coset space measurable with respect to the Haar 
measure on the group \SLs is too small: it does not contain the 
constants. Furthermore: the coset space \scosets is noncompact. 
This problem is solved by exhausting the coset space by 
a sequence of compact subsets. The restriction of the solutions 
on each compact subset of \scosets form (by group theoretical 
arguments) the required subalgebra of the Stone-Weierstrass 
theorem. The sequence of compact subsets converges towards 
\scoset. 

The Hilbert space ${\cal H}$ is defined to be the closure 
of the space of solutions of the eigenvalue equations corresonding 
to $\la, \wi{J^+}$ or $\la, \wi{J^0}$, respectively. 
It turns out that the K-Bessel functions (\ref{bes2}b) 
constitute a basis in the case of the diagonalization of 
$\la, \wi{J^+}$ and the Hilbert space is the closure of the span of 
these functions. For the diagonalization of $\la, \wi{J^0}$ the 
functions (\ref{le1}a) provide a basis. It can be shown that the 
Hilbert spaces are isomorphic.   
The scalar product is defined employing the Haar  measure on 
the coset space dressed with a suitable damping factor,  
which is chosen such that the constants and enough functions 
to separate the points on \scosets belong to the Hilbert space.
That is, from a technical point of view  the Hilbert space theory 
of the continuation of symmetric operators to their closure is 
applied. 

One realization of the coset space \scosets is the Poincar\'{e} 
upper half plane \HH. The isomorphism is established  now \cite{Lan75}. 
Each element $g$ of the group \SLs can be represented by 
a $2 \times 2$ matrix with real entries and determinant 
equal to one.  
\[
   \SL  := \left\{ g = \left( \ba{cc}
                      a & b \\
                      c & d
                    \ea \right), \quad ad - bc = 1 \right\}   
\]
\HHs is defined to be  
\[
  \HHs := \{ z = \go + i \: \tau \in \C ,~ \tau > 0\}. 
\]
Let the \SLs act on \HHs in the following manner:
\[
  g \:\circ\: z \:=\: \frac{a z \: +\: b}{c z \:+\: d}.  
\]
It is easy to see, that the mapping $g \ms g\:\circ\:i$ 
from \SLs into \HHs induces a bijection \scosets $\ra \HH$. 
  
As motivated above, the Hilbert space ${\cal H}(\HH, d\mu_n) $ 
is the space of square integrable functions 
on the upper half plane, which are measurable with respect to 
the \SLs \\ invariant measure $d\mu$   
\[ 
  d\mu \:=\: \frac{d\go d\tau}{\tau^2}. 
\]
dressed with the damping factor 
\[
    d_n(\go, \tau) \:=\: \frac{\tau^n}{
    \left( \go^2 \:+\: (1 + \tau)^2 \right)^n}, \qquad \in \N.      
\]
The Hilbert space measure is defined by $d\mu_n = d_n(\go, \tau) d\mu$. 
It is shown below, that ${\cal H}(\HH, d\mu_n)$ contains solutions 
of the differential equations as long as $n > 1$. Then they are a 
dense subset of ${\cal H}(\HH, d\mu_n)$.   
With $d_{n_1}(\go, \tau) \:<\: d_{n_2}(\go, \tau)$ for
$n_1 \:>\: n_2$ there exists a natural embedding of the $L^2$
spaces with a smaller $n$ into those with a greater $n$.
In particular ${\cal H}(\HH, d\mu_0) = L^2(\HH).$
The scalar product $(\:,\:)_n$ is defined by
\[
  (\: f(\go, \tau), \: g(\go, \tau))_n \:=\:
    \int_{-\infty}^\infty \!\!\!\!\! d\go
    \int_0^\infty  \!\!\!\!\! d\tau\:
    \frac{d_n(\go, \tau)}{\tau^2} \: f(\go, \tau) g^\star(\go, \tau) \:
\]
Concerning (\ref{dgl1}) it is found 
\begin{lemma}
  For $1 - 4 \gl \:>\:0$ neither $\psi_{\gl L}^1$ nor $\psi_{\gl L}^2$
  belong to ${\cal H}({\rm \HH}, d\mu_n)$. \\
  For $1 - 4 \gl \:\le\: 0$ the $\psi_{\gl L}^2$ are elements of
  ${\cal H}({\rm \HH}, d\mu_n)$, $n > 1$.
\end{lemma}

It is sufficient to show that
\[
  (\: \psi^1_{\gl L}, \: \psi^1_{\gl L})_n \:=\:
    \int_{-\infty}^\infty \!\!\!\!\! d\go
    \int_0^\infty  \!\!\!\!\! d\tau\:
    \frac{d_n(\go, \tau)}{\tau^2} \: |\psi^1_{\gl L}(\go, \tau)|^2  \:
\]
and $(\: \psi^2_{\gl L}, \: \psi^2_{\gl L})_n$ is finite.
This is essentially proven by splitting the integrals
into three parts in $\tau$, namely the asymptotic region
near zero, a singularity-free part having ``compact support''
and the asymptotic region near infinity.

At first the case $1-4\gl \:>\:0$ is studied more closely.
Near infinity 
$|\: \mbox{I}_\gk(|L| \tau)\:| \:\sim\: \me^{\tau} $  holds and
therefore the integral does not exist. On the other hand
$|\: \mbox{K}_k(|L| \tau)\:| \:\ge\: k_0\: \tau^{-\frac{k}{2}} $,
which lets the norm of $\psi_{\gl L}^2$ tend to infinity, too.

For $1-4\gl \:=\: 0$ it is clear, that $I_0$ is not bounded
and therefore $\psi_{\gl L}^1$ is not either.
Near zero one can approximate K$_0(x)$ by $\ln \frac{2}{x}$
and near infinity by $\sqrt{\frac{\pi}{2x}} \me^{-x}$. Splitting
the integrals into the three parts mentioned above it can be
shown that the norm of $\psi_{\gl L}^2$ is finite. That
the integral over $\psi_{\gl L}^2$ remains finite for $1 - 4 \gl <0$, too,
is due to the fact that K$_0(|L| \tau)$ can be used as an upper
bound for the function K$_k$.
The easiest way to see this is to use the integral representation
for K$_k$.

To prove the divergence of
$(\: \psi_{\gl L}^1, \: \psi_{\gl L}^1)$ for $1 - 4 \gl \:<\:0$
the asymptotic expansion
\[
  \mbox{I}_\nu(z) \:=\: \frac{\me^z }{ \sqrt{2 \pi z}} \:
  \sum_{j=0}^\infty \: (-)^j
  \frac{(4 \nu^2 -1) (4\nu^2 -3) ...
    [4 \nu^2 - (2j-1)^2]}{8^j \: j! \: z^j}
\]
is employed.

Now one answers the question the norm of which
part of the solutions $\psi_{\gl M}$ of the system of 
differential equations (\ref{dgl2}) is finite. The parameter $\gk$
assumes values on the positive real line $\R_+$ or on
the positive imaginary line $ i \R_+$.
\begin{lemma}
  For $1 - 4 \gl > 0$  and for $1 - 4 \gl \le 0$ the $\psi_{\gl M}^1$
  belongs to ${\cal H}({\rm \HH}, d\mu_n), \: n>1$.
\end{lemma}
This lemma is proven by rather tricky estimates of the integrals
$(\: \psi_{\gl M}^1, \: \psi_{\gl M}^1)_n$. The first
integral is reduced to one over the damping factor $d_n$
using that by holomorphy there exists an upper bound for
the Hypergeometric function on the interval $[0,\: 1/2]$.
Concerning the second integral the existence of a lower
bound in the neighbourhood of 0 is quite useful to prove
divergence of the integral.

The Hilbert space is spanned by the solutions of the systems of 
differential equations (\ref{dgl1}) or (\ref{dgl2}), respectively. 
In particular they form a dense subspace of the Hilbert space. 
Therefore the symmetry has to be proven on the solutions of these 
differential equations only where the operators are diagonal. 
That is, by construction of the Hilbert space the current operators 
$\wi{J^0}, \wi{J^+}$ and the Casimir operator $\la$ are already 
symmetric. The self-adjointness is discussed with the 
technique of the deficiency indices $n_+$ and $n_-$.
Given a symmetric operator $A$, the deficiency index $n_\pm$ is
defined to be the dimension of the space $\mbox{ker}( i \pm A^\star)$.
The operator is essentially self-adjoint iff the deficiency 
indices are zeroi. It has a self-adjoint extension, 
iff the deficiency indices are equal. Therefore the dimensions  
$n_-$ of the kernel of $i - \wi{J^+}^\star$ and $n_+$ of 
$i + \wi{J^+}^\star$ are calculated next.
\begin{lemma}
  The deficiency indices $n_+$ and $n_-$ are equal to zero, i.e.
  the operator $\wi{J^+}$ is already essentially self-adjoint in 
  ${\cal H}({\rm \HH}, d\mu_n)$.
\end{lemma}
The kernel, the dimension $n_+$ of which is calculated, consists
of the functions $\psi_{\gl L}$ with $\gl \:=\: i$ and $L \:=\: -i$
and the ones corresponding to $n_-$ consists of $\psi_{\gl L}$ with
$\gl \:=\: -i$ and $L \:=\: i$. The techniques to prove the
preceeding Lemma are applied to show the divergence of all
the integrals involved.

In analogous manner one calculates the dimensions
$n_-$ of the kernel and $i - \wi{J^0}$ and $n_+$ of
$i + \wi{J^0}^\star$.
\begin{lemma}
  The deficiency indices $n_+$ and $n_-$ are equal to zero.
  That is the operator $\wi{J^0}$ is already essentially 
  self-adjoint in ${\cal H}({\rm \HH}, d\mu_n)$.
\end{lemma}

$\la$ is shown to be essentially self-adjoint, too. With the 
solutions of the differential equations (\ref{dgl1}, \ref{dgl2}) 
it follows, that  
\begin{lemma}
  The spectrum of the operators $\wi{J^0}$ and $\wi{J^+}$ 
  consists of the whole real line.
\end{lemma}

To summarize: part of the formal simultaneous eigendistributions 
of the invariant differential operator $\la$ and the
Taub-NUT operator $\wi{J^+}$ as well as of $\la$ and the
physical ``mass-operator'' $\wi{J^0}$ are shown to belong to
${\cal H}(\HH, d\mu_n)$. They are exactly the functions
which are needed to decompose a function on the Poincar\'{e}
upper half plane into irreducible parts.
This is dicussed in the group theoretical context
below. The operators $\wi{J}^+$ and $\wi{J}^0$ are shown to be
essentially self-adjoint. Finally no constraints on the
spectra of the operators $\wi{J^+}$ and $\wi{J^0}$ arise:
they consist of the whole real line. Negative Taub-NUT charges
as well as negative Schwarzschild masses belong to the spectrum, too.

Now one switches to the group theoretical point of view.
There exists a canonical Hilbert
space associated to the group \SL. The unitary irreducible
characters provide a basis of this Hilbert
space. There are sufficiently many of them to separate the points of
the group which means that a function on the group can be
approximated by its characters: one obtains a generalized
Fourier transform. 
This is the reason why the construction of the Hilbert space 
above makes sense. The property that the characters separate 
the points of the group reminds one of the assumption of the 
Stone-Weierstrass theorem mentioned above. 
The generalised inverse Fourier transform, which is called 
Plancherel theorem in the mathematical literature, yields an 
expansion of an arbitrary function on the group as a series 
or an integral of functions which occur as matrix elements 
of unitary irreducible representations on the group. 
It contains the characters of the continuous and the discrete
series. 
The formula can then be applied to elements of \scosets  
with the result, that only one part of the continuous series 
survives. The eigendistributions 
of the invariant differential operators are shown to appear 
explicitely in the inverse Fourier transform. 
That is the deeper reason why the spectra of the
mass and the charge operator are purely continuous.

The classification of the unitary irreducible representations
of \SL is due to Bargman \cite{Bar47}: 
\begin{enumerate}
  \item the principal continuous series $\: V^{j,s}, 
    \quad j = 0, 1/2, \: s = 1/2 + it,\: t \in \R.$
    If $j = 0$, then $t > 0$, and if $j = \frac{1}{2}$, then
    $t \ge 0$.  
  \item the limit of the discrete series $\: U^{1/2}, 
    \: U^{-1/2}$, 
  \item the discrete series $\: U^n, \: U^{-n}, \: n \in \Z/2, 
    \: n > 1/2$, 
  \item the complementary series $\: V^\gs, \: 1/2 < \gs < 1 $ and
  \item the trivial representation. 
\end{enumerate}
Let $\gl$ denote the eigenvalue
of the Casimir operator. To the five cases listed above
there correspond the following eigenvalues: 
\begin{enumerate}
  \item $\gl = s(1-s),$ 
  \item $\gl = \frac{1}{4},$
  \item $ \gl = n(1-n),$
  \item $ \gl = \gs ( 1 - \gs )$, and 
  \item $\gl = 0$. 
\end{enumerate}
That is, group theory yields a restriction of the eigenvalues of
the Casimir operator and some formal solution of the differential
equations are excluded from the Hilbert space by group theoretical
arguments, namely the case $1 - 4 \gl >0$ mentioned above.
To the representations there correspond characters, which
are class functions. To evaluate and to apply
them it is quite natural to factor the group by the Iwasawa
decomposition into an elliptic,
a hyperbolic and a parabolic part, which are denoted by
$G_{_{\rm ell}}, \: G_{_{\rm hyp}}$ and $G_{_{\rm par}}$, 
respectively. Now it is outlined how to proceed: 

Each $2 \times 2$ matrix of \SLs can be written as a product of  
a rotation matrix, the elliptic part of \SL, a diagonal matrix 
with entries either both positive or both negative, the hyperbolic 
part of the group,  and one matrix with a Jacobi-like upper 
triangular form, which parametrizes the parabolic part of the group. 
\bes
  \SL &\:=\:& K \: A \: N \\
   g  &\:=\:& u_{_\gt} \: a_\tau \: \eta_\go  
\ees
with
\[
   u_{_\gt}  = \left( \ba{cc}
                      \cos \frac{\gt}{2} & \sin \frac{\gt}{2} \\[1.5ex]
                      -\sin \frac{\gt}{2} &\cos \frac{\gt}{2}  
                    \ea \right), \qquad    
   a_\tau  = \left( \ba{cc}
                      \tau & 0 \\
                      0 & \tau^{-1} 
                    \ea \right), \qquad    
   \eta_\go  = \left( \ba{cc}
                      1 & \go \\
                      0 & 1 
                    \ea \right)    
\]
Denote $H \: :=\: A \:\cup\: (-1) \: A$, then 
$G_{_{\rm hyp}}, G_{_{\rm ell}}$ and  $G_{_{\rm par}}$ are defined by 
\[
  G_{_{\rm hyp}} \: := \: \:\bigcup_{g \in G}\quad g \: H \: g^{-1}, \quad
  G_{_{\rm ell}} \: := \: \:\bigcup_{g \in G}\quad  g \: K \: g^{-1}, \quad
  G_{_{\rm par}} \: := \: \:\bigcup_{g \in G}\quad g \: (\pm N) \: g^{-1}.
\]
Furthermore the group \SLs can be decomposed into disjoint 
parts of classes of regular elements, i.e. elements with 
distinct eigenvalues, 
$G_{_{\rm ell}}^\prime,\: G_{_{\rm hyp}}^\prime$ and 
$\: G_{_{\rm par}}^\prime$. $\: G_{_{\rm par}}^\prime$ is the empty set.  
$G$ may be represented as the product of $G_{_{\rm ell}}^\prime, \: 
G_{_{\rm hyp}}^\prime$ and $G_{_{\rm par}}$. 
The series of the unitary irreducible representations correspond 
in a one to one manner to the factors of the Iwasawa decomposition:
the discrete series to the elliptic part, the continuous one to the 
hyperbolic factor and the complementary series to the parabolic one.
An algorithm to compute the Iwasawa decomposition for elements of 
one of the classical Lie groups can be found in the {\bf Appendix A}.

Before dealing with the decomposition of a function on \SLs into 
its irreducible parts, the space of rapidly decreasing functions 
${\cal S}(G), G = \SL$ has to be introduced (see \cite{BarRac77}):
\[
  {\cal S}(G) \:  := \: \{ f \in C^\infty(G): \quad
     \sup_{g \in G} \frac{\left( \go^2 + (1 + \tau)^2\right)^N}{\tau^N} 
     \: |(D^\ga f)(g)| \: < \: \infty, \quad \forall N \}
\]
The tempered distributions ${\cal S}^\prime(G)$ are the dual space 
of the space of the rapidly decreasing functions. They are called 
slowly increasing. The triplet ${\cal S}(G), L^2(G, d\mu_G)$ 
and ${\cal S}^\prime(G)$ is called a Gel'fand triplet. $d\mu_G$ 
denotes the Haar measure on \SL.
It holds 
\[
   {\cal S}(G) \quad \subset \quad L^2(G, d\mu_G) \quad 
   \subset \quad {\cal S}^\prime(G)
\]
As already mentioned above, ${\cal H}(G, d\mu_G) = L^2(G, d\mu_G)$. 
${\cal S}^\prime(G)$ contains all of the spaces ${\cal H}(G, d\mu_n)$.
${\cal H}(G, d\mu_G)$ denotes the Hilbert space of functions on $G$, 
which are measurable with respect to the Haar measure of $G$.  

This approach is similar to the one needed for the investigation 
of the Hydrogen Atom. There the coset space SO(3)/SO(2) which is 
isomorphic to the 2-sphere $S^2$ plays the key role. The generalised 
Fourier transform of SO(3) represented on $S^2$ leads to the 
Spherical Harmonics. The main difference arises from the fact, 
that the group \SLs is no longer compact, therefore the unitary 
irreducible representations are no longer finite dimensional,  
and apart from a discrete sum there further appears a direct 
integral in the decomposition formula. 

The generalized Fourier transform \cite{Sug90} establishes a 
topological isomorphism between the $C^\infty$ functions  
the space of rapidly decreasing function on  
the dual of the group\footnote{$\hat{G}$ is defined to be 
the set of all equivalence classes of unitary irreducible 
representations of the group.} $\hat{G}$.
\begin{theorem}
  For each function $f \in {\cal S}(G)$
\[
  f(g) \quad=\quad \frac{1}{2 \pi} \: \int_0^\infty \!\!\! dt \:
        \mbox{\rm Tr}[ \hat{f} (0, \frac{1}{2} + it) \:
     V_g^{0, \frac{1}{2} + it} ] \quad t \: \tanh \: \pi t
\]
\[
       \quad+\quad \frac{1}{2 \pi} \: \int_0^\infty \!\!\! dt \:
        \mbox{\rm Tr}[ \hat{f} (\frac{1}{2}, \frac{1}{2} + it) \:
     V_g^{\frac{1}{2}, \frac{1}{2} + it} ] \quad t \: \coth \pi t
\]
\be
\label{plan}  
     + \frac{1}{4 \pi} \:
     \sum_{\stackrel{n \in \frac{1}{2} \Z}{n \ge 1}} \:
     ( 2n -1) \: \left\{ \mbox{\rm Tr} \:[ \hat{f}(n)\: U_g^n ]
       \:+\: \mbox{\rm Tr} \: [\hat{f}(-n)\:U_g^{-n} ] \right\}.
\ee
   In particular:
\[
  f(1) \quad=\quad \frac{1}{2 \pi} \: \int_0^\infty \!\!\! dt \:
     \left(
        \gT^{0, \frac{1}{2} + it}(f) \: t \: \tanh\: \pi t
     \quad+\quad
       \gT^{\frac{1}{2}, \frac{1}{2} + it}(f) \: t \: \coth\: \pi t 
    \right)
\]
\[
    + \quad \frac{1}{4 \pi} \:
    \sum_{\stackrel{n \in \frac{1}{2} \Z}{n \ge 1}}
   \: ( 2n -1) \: [ \gT^n(f) \:+\: \gT^{-n}(f)]
\]
with
\[
  \Theta^{0, \frac{1}{2} + it} (f) \:=\: \int_G \!\!\! dg \:
  f(g^{-1}) \Theta^{0, \frac{1}{2} + it} (g)
\]
\end{theorem}
$\gT^{0, \frac{1}{2} + it}$ and $\gT^{\frac{1}{2}, \frac{1}{2} + it}$ 
denote the characters of the principal continuous series and $\gT^n$ 
and $\gT^{-n}$ those of the discrete series. Only the discrete and the 
principal continuous series contribute to the support of the Plancherel 
measure in $\hat{G}$ in the Plancherel theorem for  a general group 
element $g$ of \SL. 

In Einstein's gravity a parametrization of the coset space is 
given: the upper half plane. Therefore the formula has to be 
applied to \HH. 
One starts with a function $f \in {\cal S}(\HH)$, which is  
invariant under the action of the elliptic part $K$ of the 
group. This leads to the conclusion that the irreducible characters 
associated with $K$ do not contribute to the Fourier inversion 
formula for such functions.  
Without loss of generality $g$ can be taken to be 
an regular element of the hyperbolic part of the group 
$G_{_{\rm hyp}}^\prime$. One calculates 
\[
  \mbox{\rm Tr} \hat{f}(n) \: U_g^n \:=\:
   \mbox{\rm Tr} \left( \int_G \!\!\! dl \: f_g(l) \: U_{l^{-1}}^n \right).
\]
$f$ and $f_g$ are both left and right invariant under the action of $K$. 
The representation 
$U_{l^{-1}}^n$ is explicitely known. It acts on holomorphic 
functions $s$ on \HHs in the following manner [Bar47]: 
\[
  U_{l^{-1}}^n s(z) \:=\: \frac{1}{(cz +a )^{2n}} \:
  s \left( \frac{az + b}{cz + d} \right), \qquad 
  l^{-1} \:=\: 
  \left(
    \ba{cc}
      a & b \\
      c & d 
    \ea
  \right).
\]
for a rotation matrix $l$ the action reads   
\[
  U_{l^{-1}}^n s(z) \:=\: \frac{1}{
  (\sin\frac{\gt}{2} z + \cos \frac{\gt}{2} )^{2n}} \:
  s \left( \frac{ \cos \frac{\gt}{2} z - \sin \frac{\gt}{2} }{
  \sin \frac{\gt}{2} z + \cos \frac{\gt}{2} } \right).
\]
The integrand is a holomorphic function on \HHs and the path 
of integration is closed. 
Therefore the integral vanishes and the discrete series does not 
contribute to the Plancherel formula on \HH. 
The characters of the continuous series enter into the formula 
in two places. 

Due to the symmetry the $\coth$-part denoted by $f_2(g)$ does 
not contribute to the formula either. 
By translation all group elements are obtained from unity and 
\be
\label{trans}
  \mbox{\rm Tr}
 \left[
   \hat{f}(\frac{1}{2}, \frac{1}{2} + it) \:
   V_g^{\frac{1}{2}, \frac{1}{2} + it}
  \right] \:=\: \gT^{\frac{1}{2}, \frac{1}{2} + it}(f_g)      
\ee
$f_g$ is two-sided $K$-invariant leading to $f_g(g) \:=\: f_g(-g)$. 
The character $\gT^{\frac{1}{2}, \frac{1}{2} + it}(g)$ is 
\cite{Sug90} 
\[
  \gT^{\frac{1}{2},\frac{1}{2} + it}(g) \: = \:
  \left\{
    \ba{ll}
     \displaystyle{\frac{\cos t \tau}{ |\sinh \tau |}}, \: &
     g \: = \: g_0 a_\tau g_0^{-1} \: \in \: G_{_{hyp}}^\prime \\
     & \\
     -\displaystyle{\frac{\cos t \tau}{ |\sinh \tau |}}, \: &
     g \: = \: - g_0 a_\tau g_0^{-1} \: \in \: G_{_{hyp}}^\prime \\
     & \\
     0, & g \in G^\prime_{_{ell}} \cup G_{_{par}}
     \ea \right.
\]
It follows that $\gT^{\frac{1}{2}, \frac{1}{2} + it}(g) \:=\: 
\gT^{\frac{1}{2}, \frac{1}{2} + it}(-g)$, i.e. the character is 
point symmetric in $\tau$ and (\ref{trans}) vanishes. That is, 
the coth-part does not contribute to the Plancherel theorem on 
the upper half plane, either. 

Given a representation space, which is the Poincar\'{e} upper half 
plane here, it is now possible to prove, that the Plancherel 
theorem on \scosets can be written such that the eigendistributions of 
the Casimir operator appear as integrand in the formula!
\begin{theorem}
  For $f \in {\cal S}({\rm \HH})$, 
  \[
     f(\go,\tau) \:=\: \frac{1}{\pi^2} \:
     \int_{L \in \R} \!\!\!\!\! dL \int_{t \in \R} \!\!\!\!\! dt
      \: \hat{f}(L,t) e_{L,t}(\go,\tau) \: t \sinh\: \pi t
  \]
  where
  \[
    e_{L,t}(\go,\tau) \: = 
       \exp(- i L \go) \:\sqrt{\tau} \: K_{it}(|L| \tau),
         \quad  \mbox{if} \quad L \neq 0 
  \]
  and
  \[
     \hat{f}(L,t) \: =\: \int_{\rm \HH} \!\!\! d\go d\tau \:
     \frac{f(\go,\tau) \: \overline{e_{L,t}(\go,\tau)}}{\tau^2}
  \]
\end{theorem}
{\bf Proof:}\\
It is sufficient to show, that the tanh-part of the Plancherel 
formula (\ref{plan}), denoted by $f_1(g)$ is equivalent to this 
lemma.  The nuclear spectral theorem \cite{BarRac77} yields 
\bea 
\label{ft}
  & & f(\go, \tau) \:=\: 2 \pi \: 
    \int_{L \in \R} \!\!\! dL \int \!\!\! d\gl(t) \: \hat{f}(L,t) 
    e_{L,t}(\go, \tau)  \\
  &=& 2 \pi \:
      \int \!\!\! d\gl(t) \int_{\tilde{\go} \in \R} \!\!\! d\tilde{\go} 
      \int_{\tilde{\tau} \in \R_+} \!\!\! d\tilde{\tau} \:
      \frac{f(\tilde{\go}, \tilde{\tau})}{\tilde{\tau}^2} \:
    \left( 
      \int_{L \in \R} \!\!\! dL \:    
      \overline{e_{L,t}(\tilde{\go}, \tilde{\tau})} \:
      e_{L,t}(\go, \tau)
    \right)  \nonumber 
\eea
The inner integral can be evaluated using the calculus of residues, 
which leads to an orthogonality relation for the K-Bessel functions:
\be
\label{otog}
   \int_{L \in \R} \!\!\! dL \:    
   \overline{e_{L,t}(\tilde{\go}, \tilde{\tau})} \: 
   e_{L,t}(\go, \tau) 
   \:=\: \gd(\go - \tilde{\go}) \: \gd(\tau - \tilde{\tau}) \: 
   \frac{\pi}{4 \cosh t \pi}
\ee
Substitution of (\ref{otog}) into (\ref{ft}) yields 
\be
  \label{pt1}
  f(\go, \tau) \:=\: \frac{\pi^2}{2} \:
  \int_{t \in \R} \!\!\! d\gl(t) \int_{\tilde{\go} \in \R} 
  \!\!\! d\tilde{\go} 
  \: \frac{f(\tilde{\go}, \tilde{\tau})}{\tilde{\tau}^2} \: 
  \gd(\go - \tilde{\go}) \: \gd(\tau - \tilde{\tau}) \: 
  \frac{1}{\coth \pi t}.
\ee 
On the other hand $f_1(g) \:=\: f_{1,g}(1)$. Therefore and because 
\SLs is unimodular $f_1(g)$ is transformed into 
\[
  f_1(g) \:=\: \frac{1}{2 \pi} \: \int_0^\infty \!\!\! dt \:
    \gT^{0, \frac{1}{2} + it}(f_g) \: t \: \tanh \pi t 
    \:=\: \frac{1}{2 \pi} \: 
    \int_0^\infty \!\!\! dt \int_G \!\!\! d\tilde{g} \: 
    f_g(\tilde{g}^{-1}) \: \gT^{0, \frac{1}{2} + it}(\tilde{g}) \: 
    t \: \tanh \pi t
\] 
The character is symmetric with respect to $t$ [Sug90] and the 
\SL acts on the upper half plane leading to 
\bea
  \label{pt2}
  f_1(h) &=& \frac{1}{4 \pi} \: 
  \int_{-\infty}^\infty \!\!\! dt \int_{\HH} \!\!\! d\tilde{h} \: 
  f_h(\tilde{h}^{-1}) \: \gT^{0, \frac{1}{2} + it}(\tilde{h}) \: 
  t \: \tanh \pi t  \nonumber \\
  &=& \frac{1}{4 \pi} \:
  \int_{-\infty}^\infty \!\!\! dt \int_{\HH} \!\!\! d\tilde{h} \: 
  f_h(\tilde{h}) \: \gT^{0, \frac{1}{2} + it}(\tilde{h}^{-1}) \: 
  t \: \tanh \pi t  \nonumber \\
  &=& \frac{1}{4 \pi} \:
  \int_{-\infty}^\infty \!\!\! dt \int_{\HH} \!\!\! d\tilde{h} \: 
  f(\tilde{h}) \: \gT^{0, \frac{1}{2} + it}_h(\tilde{h}^{-1}) \: 
  t \: \tanh \pi t  
\eea
The normed character ($\Phi^{0, \frac{1}{2} + it}(e) = 1$) is 
\[
  \gT^{0,\frac{1}{2} + it}(g) \: = \:
  \left\{
    \ba{ll}
     \cos t \tau, \: &
     g \: = \: g_0 a_\tau g_0^{-1} \: \in \: G_{_{hyp}}^\prime \\
     & \\
     0, & \mbox{otherwise}
     \ea \right.
\]
(\ref{pt1}) and (\ref{pt2}) yield 
\[
  \gl^\prime(t) = \frac{1}{4 \pi} \: 
  \gT^{0, \frac{1}{2} + it}(e)\: t \: \tanh \pi t \: \frac{2}{\pi^2} 
  \cosh \pi t = \frac{t \sinh \pi t}{2 \pi^3} \: 
  \gT^{0, \frac{1}{2} + it}(e)
\]
and therefore 
\[
  \gl^\prime(t) \:=\: \frac{1}{2 \pi^3} \: t \: \sinh \pi t
\]
This concludes the proof. 

\hspace*{13cm} $\Box$

This Fourier inversion formula corresponds to a simultaneous 
diagonalization of the Casimir operator and the differential operator 
corresponding to the generator of the Lie algebra $\sl$, which has 
a 1 in the upper right corner and 0's elsewhere. The Taub-NUT 
charge $L$ appears explicitely. On the algebraic level $\wi{J^0}$ 
belongs to the generator diag$(1, -1)$. 

The representation most 
suitable for the Schwarzschild mass $M$ yields another Plancherel 
inversion formula. Given a  $f \in {\cal S}({\rm \HH})$, 
this function can be written as  
\[
   f(\go,v) \:=\: C \:
   \int_{M \in \R} \!\!\!\!\! dM \int_{t \in \R} \!\!\!\!\! dt
    \: \hat{f}(M,t) g_{M,t}(\go,v) \: d \gl(M,t) 
\]
where
\[
  g_{M,t}(\go,\tau) \: =
  \go^{i M} \: \sqrt{v} 
   \left( 1 + v^2\right)^{i \frac{M}{2} - \frac{1}{4}} \:
   \mbox{P}^{-\frac{\gk}{2}}_{-i M  -\frac{1}{2}} 
  \left( \frac{1}{\sqrt{1 + v^2}}\right),
        \mbox{if} \quad M \neq 0 \\
\]
and
\[
   \hat{f}(M,t) \: =\: \int_{\rm \HH} \!\!\! d\go dv
   \: \frac{f(\go,v) \: \overline{g_{M,t}(\go,v)}}{v^2 \go}.
\]

After the diagonalization of the Casimir operator the 
Wheeler-DeWitt equations is transformed into
\be
  \label{wdws}
  \psi^{\prime \prime} + ( 4 - \frac{2 \gl}{f^2} ) \psi = 0.
\ee
$\gl$ denotes the eigenvalue of the Casimir operator. 
(\ref{wdws}) can be interpreted as evolution equation in the 
coordinate $f$. In contrast to $\rho$, $f$ has a physical meaning 
as $\frac{1}{f}$ is the curvature of the 2-spheres. Hence it is 
also a physical parameter. Equation (\ref{wdws}) is a second order 
differential equation and therefore there does not exist a 
positive semi-definite probability density which is invariant under 
$\SLs$ transformations and conserved during evolution. 
This situation is comparable with the one in the case of the 
Klein-Gordon equation: a conserved current can be associated 
to the Klein-Gordon equation, but the zero component of the 
current is not positive definite. Feshbach and Villar \cite{FesVil58}  
interpreted $j_0$ as a charge density which measures  
the difference between the numbers of positive and negative 
charges. In the one-particle case, they showed that the density 
carries either a positive or a negative sign and that the two degrees of 
freedom of the second order differential equation are identified 
as two possible but equivalent charge states. In \cite{FesVil58} it 
is outlined that although the nonrelativistic equation only 
admits charges of one sign and in the relativistic generalization 
two signs of the charge happen to appear. 
The two equivalent charge states are basically obtained by 
transforming the Klein-Gordon equation into a symmetrized 
Schr\"{o}dinger equation, i.e. a system of first order 
differential equations. 
Now equation (\ref{wdws}) is investigated in two steps. 
At first, the summand $-\frac{2 \gl}{f^2} \psi$ is 
considered to be a $f$-dependent perturbation and is neglected. 
This ``free'' equation can easily be solved and two equivalent 
states are found. The second step consists in investigation 
of the ``perturbed'' system. With a given and finite $\gl$ the 
solution approaches a the free solution in the limit $f \ra \infty$.  
This fixes the asymptotic behaviour of the solution. 
The two linear independent solutions can then again be shown 
to form two equivalent but independent states. 

The ``free'' differential equation $\psi_0^{\prime \prime}$ 
has the solution 
$\psi_0(f) = C_1 \mbox{e}^{2 i f} + C_2 \mbox{e}^{-2 i f}$. 
It follows that 
\[
  \psi_0^\star \pa_f \psi_0 - \psi_0 \pa_f \psi_0^\star 
  = \frac{i}{4} ( \chi_0^\star \chi_0 - \phi_0^\star \psi_0 )
  = \frac{i}{4} (C_1^2 - C_2^2 )
\]
with $\phi_0 = C_1 \mbox{e}^{2 i f}$ and 
$\chi_0 = C_2 \mbox{e}^{-2 i f}$. It is not difficult to 
solve the differential equation (\ref{wdws}). The solutions 
with the proper asymptotic behaviour turn out to be the 
Bessel functions of third kind $H^{(1)}_\nu$ and $H^{(2)}_\nu$:
\[
  \psi(f) = C_1 \sqrt{f} H^{(1)}_\nu (2 f) 
    + C_2 \sqrt{f} H^{(2)}_\nu(2 f), 
  \qquad \nu = \frac{1}{2} \sqrt{1 + 8 \gl}.   
\]
Using that the complex conjugate of $H^{(1)}_\nu$ is $H^{(2)}_\nu$, 
the density is calculated to be 
\bea
  & & \psi \pa_f \psi - \psi \pa_f \psi \\ \nonumber 
  &=& 2 C_1^2 f 
  \left[
    H^{(2)}_\nu ( 2f) \pa_f H^{(1)}_\nu(2f)  
  - H^{(1)}_\nu ( 2f) \pa_f H^{(2)}_\nu(2f)  
  \right]  \\ \nonumber 
  &+& 2 C_2^2 f
  \left[
    H^{(1)}_\nu ( 2f) \pa_f H^{(2)}_\nu(2f)  
  - H^{(2)}_\nu ( 2f) \pa_f H^{(1)}_\nu(2f)  
  \right] \\ \nonumber
  & = & \frac{8 i }{\pi} ( C_1^2 - C_2^2 ).
\eea
The last equaltity uses the Wronskian of $H^{(1)}_\nu $ and 
$H^{(2)}_\nu$.
Two ``equivalent'' states in this case are 
\[  
  \phi = \frac{1}{2} 
  \left( \pa_f \psi + i \psi \right), 
  \qquad \mbox{and} \qquad
  \chi = \frac{1}{2} 
  \left( \pa_f \psi - i \psi \right).
\]
In these new variables the density $\rho$ reads  
$\rho = 2 i (\chi \chi^\star - \phi \phi^\star).$ 

The choice of the equivalent states is not unique. Define $\chi$ 
and $\phi$ by 
\[
  \left( \ba{l}
       \psi \\
       \pa_f \psi
         \ea \right) 
  = \left( \ba{ll}
      a_{11} & a_{12} \\
      a_{21} & a_{22} 
           \ea \right) 
  \left( \ba{l}
       \chi \\
       \phi
         \ea \right).   
\]
As long as $ \bar{a}_{12} a_{21} - \bar{a}_{22} a_{11} = 0$ the 
states $\chi$ and $\phi$ are equivalent in the sense that $\rho$ 
is the difference of two positive semi-definite densities. 
One part of the freedom can be fixed by normalisation. 

Yet the physical meaning of this discrete symmetry (``charge 
conjugation'') is not clear. 

There is a nice relation
between the quantization of the four dimensional reduced spherically
symmetric gravity in its dual representation and a \SLs WZNW model
in the point particle version, which is further reduced to a
Liouville theory [Ful96]. 
There are also some important differences which lead to a deeper 
understanding of the quantization of the system dealt with in this 
paper.  
Performing the reduction to Liouville theory F\"{u}l\"{o}p ended 
with a Hamiltonian system with constraints, which correspond to 
initial values.
Choosing these initial values the Hamiltonian of the latter system
has exactly the form of the remaining part of the Wheeler-DeWitt
equation (\ref{wdws}), where the initial values take the place of
the eigenvalue of the Casimir operator. The Liouville Hamiltonian is
interpreted to be the Hamiltonian of a relativistic particle which
moves in a potential. Formally, there are three cases to be
distinguished depending on the sign of the constant $\gl$ for the
initial values. If $\gl>0$, the particle is affected by an infinitely
high potential barrier, when it travels in the negative $f$-direction. For
$\gl = 0$ the particle is free, and for $\gl < 0$ an infinitely deep
potential valley attracts the particle towards the negative $f$-direction.
This interpretation is in agreement with the group theoretical 
facts.  
$\gl > 0$ corresponds to $\gl = s ( 1 - s) = \frac{1}{4} + t^2$ 
and these are the values of the continuous series, which lead to 
scattering states. $\gl < 0$ means $\gl = n(1-n), n \in \Z/2, n > 1$,   
that is, it is an element of the discrete series of
representation theory. Bound states are obtained. 
Here $\gl \ge 0$ holds, because the Casimir operator can be shown to be the
square of the Taub-NUT charge and the Schwarzschild mass operator.
If the Casimir operator on the group would no longer be positive 
semi-definite, one would expect the discrete series of 
representations to play a role. The spectra of the physical would 
have a discrete part.  
However, in the case of stationary spherically symmetric gravity this 
argumentation is certainly not more than a consistency check: here 
the coordinate $f$ has no intrinsic group theoretical meaning in 
contrary to the situation which occurs in the Liouville model.  
It is a ``gravitational remnant''. 

Another important difference is that F\"{u}l\"{o}p considers 
the momenta of the particle to be fundamental. In particular they 
become hermitian operators. Here, as the physical meaning 
is contained in the currents, there is no reason for the momenta to
become hermitian operators. Moreover it turns out that starting
with hermitian momenta the currents are non-hermitian. On the other
hand, the conditions that $\wi{J^0}, \wi{J^+}$ and $\wi{J^-}$ are
hermitian defines these operators to be Lie derivatives.
This also fixes the operator ordering of the Laplacian and
therefore of the Hamiltonian, too.

\section{Results and Discussion}

After proper dimensional reduction the spherically symmetric 
sector of Einstein's vacuum theory can be identified with a 
\scosets sigma model coupled to a gravitational remnant which 
belongs to a gauge degree of freedom. 
The ``true'' dynamical degrees of freedom are entirely hidden in the 
sigma model. In addition to the Schwarzschild mass $m$ 
another parameter of the classical space of solutions -- the 
Taub-NUT charge $l$ -- shows up. The invariance of the Lagrangian 
under the group \SLs offers the key to quantize this part of 
gravity. In a modified Hamiltonian formalism the invariant 
differential operator on the group -- the Casimir operator -- 
appears quite naturally in the Hamiltonian constraint known as 
Wheeler-DeWitt equation. It is possible to diagonalize the Casimir 
operator and the mass operator or the Taub-NUT charge operator 
simultaneously. The part of the eigendistributions not increasing more 
rapidly then a polynomial yields a spectral decomposition of the 
Laplacian. In the case of the mass operator one finds essentially 
associated Legendre polynomials, and in the case of the charge 
operator there appear $K$-Bessel functions. It was emphasized that 
the parametrization of the fields or in other words the chosen 
representation space plays quite an important role. 
The Plancherel formula of \SLs which allows to decompose each 
function on the group into irreducible parts can be applied to 
the coset space \scoset. In the case of group \SLs 
the continuous and the discrete series support the Plancherel 
measure whereas on the coset space one is left with one 
part of the continuous series.  
In the literature were noted some difficulties with the 
self-adjointness of the observables. This may be due to the 
restriction to the quantization of the Schwarzschild solution 
only, which corresponds to a diagonal matrix $\chi$. 
Then one is forced to use the invariant measure of $\R_+$, 
instead of the Plancherel measure on \SL. Probably 
the results of this paper indicate that sectors of gravity 
can only be quantized consistently, if one takes into account 
the whole classical space of solutions belonging to it. 

The group theoretical methods are not limited to the situation 
here. They are applicable to various other models in 
conformal field theory, string theory, quantum cosmology and 
supergravity. A list of models can be found in \cite{BreMaiGib88}.

\section{Appendix A: Some remarks about the Iwasawa decomposition  
of a general group $G$}

For an arbitrary group element of one of the classical Lie groups, which
is written as a matrix $M$, with $\det M \neq 0$, there exists an
algorithm for the Iwasawa decomposition. Two numerical methods can be
used, the Householder transformation or the Gram-Schmidt
orthonormalisation  method. Basically they work as follows \cite{Sto89}:
Multiplying $M$ from the left by a unitary matrix $Q$, an upper
triangular matrix $R$
\[
  R = \left(
  \ba{lll}
    r_{11} & \cdots & r_{1n} \\
     & \ddots & \vdots\\
    0 & & r_{nn}
  \ea \right)
\]
is obtained ($QR$ decomposition).
$R$ is further decomposed into a diagonal matrix $A$ and a strictly
upper triangular matrix $N$ with ones on the diagonal:
\[
  A = \left(
  \ba{lll}
    r_{11} &  & 0 \\
     & \ddots & \\
    0 & & r_{nn}
  \ea \right) \qquad \mbox{and} \qquad
  N = \left(
  \ba{llll}
    1 & \DI{\frac{r_{12}}{r_{11}}} & \cdots 
    & \DI{\frac{r_{1n}}{r_{11}}} \\
    0 & \ddots &  & \vdots \\
    \vdots & \ddots & 1 & \DI{\frac{r_{n-1 n}}{r_{n-1 n-1}}} \\
    0 & \cdots & 0 & 1
  \ea \right).
\]
The Gram-Schmidt orthonormalisation method directly yields the factor
$K = (Q^\dagger)^\star$. In the case of the Householder transformation
one should remind that $Q$ is the product of reflections with
$\det Q = -1$. Therefore if the rank of $M$ is even one has to multiply
one row or one column by $-1$ in order to get a rotation.
For computational convenience the parametrization should be chosen
according to the factors of the Iwasawa decomposition.

\section*{Acknowledgement}\addcontentsline{toc}
{section}{Acknowledgement}

The author is indebted to P. Breitenlohner and D. Maison for
numerous discussions on the subject.



\end{document}